\documentclass{optica-article}

\journal{opticajournal} % for journals or Optica Open

\articletype{Research Article}

\usepackage{lineno}
\usepackage{subcaption}
\usepackage{ulem}
\usepackage{appendix}
\begin{document}

\title{A high-performance quantum pulse gate in thin-film lithium niobate}

\author{Silia Babel,\authormark{1,2,†,*} Alejandra Alarcón,\authormark{1,2,†} Laura Serino,\authormark{1,2} Christian Golla,\authormark{2} Laura Bollmers,\authormark{1,2} Sebastian Lengeling,\authormark{1,2}  Jiayu Yang,\authormark{3}  Bernhard Reineke,\authormark{2} Christof Eigner,\authormark{2} Benjamin Brecht,\authormark{1,2}  Marko Lončar,\authormark{3} Laura Padberg,\authormark{1,2} and Christine Silberhorn\authormark{1,2}}

\address{\authormark{1} Paderborn University, Integrated Quantum Optics, Warburger Str. 100, 33098 Paderborn, Germany\\
\authormark{2}Paderborn University,
Institute for Photonic Quantum Systems (PhoQS), Warburger Str. 100, 33098 Paderborn, Germany\\
\authormark{3} John A. Paulson School of Engineering and Applied Sciences, Harvard University, Cambridge, MA 02138, USA. \\
\authormark{†} The authors contributed equally to this work.}
\email{\authormark{*}silia.babel@upb.de} 

\begin{abstract*} 
In this work, we demonstrate a quantum pulse gate (QPG) in thin-film lithium niobate. 
QPGs enable the selective manipulation and detection of temporal modes of quantum light and form the basis of numerous applications in photonic quantum technologies. 
To date, their widespread adaption is held back by two main limitations: 
restricted wavelength and polarization combinations of the involved fields and low normalized conversion efficiencies. 
We overcome these limitations through developing a QPG in thin-film lithium niobate. 
We design a waveguide geometry that provides the required dispersion properties for a highly efficient type-0 sum-frequency generation.
We verify our design through mapping of the phase matching intensity, and demonstrate high-quality QPG operation by measuring a temporal-mode selectivity of $(96.8 \pm 1.7)\%$ on par with existing QPGs.
Thanks to the strong confinement in thin-film lithium niobate, we succeed in demonstrating an internal conversion efficiency of $(89.6 \pm 0.1)\%$ for a pump power of only $20\,\mathrm{mW}$ in front of our sample. 
This yields a lower-bound estimate for the normalized conversion efficiency of $(1810 \pm 10)\,\mathrm{W}^{-1}\mathrm{cm}^{-2}$, three orders of magnitude higher than in previous QPGs. 
Our results establish thin-film lithium niobate as ideal platform for high-performance QPGs and are a major step towards practical QPGs for photonic quantum technologies. 
\end{abstract*}

%%%%%%%%%%%%%%%%%%%%%%%%%%  body  %%%%%%%%%%%%%%%%%%%%%%%%%%
\section{Introduction}
The integration of quantum optical systems onto chip-scale platforms is a key enabler for scalable quantum technologies. 
For this to be possible in principle, information must be encoded in a basis that is compatible with integrated systems combining both photonic integrated circuits and optical fiber. 
The time-frequency degree of freedom of quantum light, more precisely the basis of temporal modes (TMs), has been recognized as an ideal candidate for fulfilling this requirement \cite{brecht2015photon,raymer20a}. 
The coherent manipulation and selection of individual TMs, a prerequisite for any quantum application, can be realized with the quantum pulse gate (QPG). 
The QPG is a TM-selective sum-frequency generation (SFG) in dispersion engineered nonlinear waveguide \cite{eckstein2011quantum, brecht2014demonstration}. 
In the past, the QPG has been used in numerous applications including fundamental quantum mechanics \cite{serino2025complementarity}, programmable unitary operations \cite{de2024realization}, quantum state tomography \cite{ansari2017temporal,serino2025self,gil2021universal, treps2019Photon-Subtracted, kowligy2014quantum}, spectral bandwidth compression \cite{allgaier2017highly}, temporal mode sorting for high-dimensional quantum communications \cite{serino2025programmable}, and quantum metrology \cite{donohue2018quantum,ansari2021achieving,namekata2023quantum,namekata2026near}. 
More recently, it has been identified as enabling tool for programmable frequency encoded quantum networks \cite{folge2024framework}. 
Realized in titanium-indiffused lithium niobate (Ti:LiNbO$_3$) waveguide, the legacy QPG combines an input at telecommunications wavelength with a strong pump at around $870\,\mathrm{nm}$ to generate a converted output in the green in a type-II SFG. 
While being highly successful in laboratory scale demonstrations, the widespread use of the QPG has been hindered by two major limitations: 
firstly, a low normalized conversion efficiency \cite{gil2021improved}  caused by the low nonlinearity of the type-II SFG and the weak confinement in the Ti:LiNbO$_3$ waveguide, which leads to unreasonable pump power requirements for high efficiency conversion; 
and secondly, severe restrictions of operating wavelengths and polarizations due to the requirement for group-velocity matching between the signal and pump, which, in weakly guiding structures, is governed by the material dispersion. 

These two limitations can be successfully overcome using thin-film lithium niobate (TFLN), which has emerged as a leading platform for integrated (quantum) photonics by enabling efficient conversion at substantially reduced pump powers. 
TFLN combines the outstanding optical properties of lithium niobate such as a high second order nonlinearity, large electro-optical coefficient and a wide transparency window \cite{weis1985lithium} with the advantages of high index contrast rib waveguides such as strong confinement, high integration density, and flexible dispersion engineering \cite{jankowski2021dispersion}. 
Numerous applications \cite{vazimali2022applications} have already been realized in TFLN, including low-loss waveguides \cite{zhu2024twenty,luke2020wafer,shams2022reduced}, electro-optic modulation \cite{xu2022dual,zhang2022systematic,hu2025integrated}, nonlinear quantum light generation and conversion \cite{zhu2023sum,wang2023quantum,sabatti2025nanodomain,kellner2025counter,zhao2020shallow, wang2026electrically} as well as quantum interference \cite{chapman2023quantum,babel2023demonstration,kuttner2025scalable,chapman2025chip}. 
Remarkably, the strong confinement enhances the effective nonlinearity of the material by several orders of magnitude compared to Ti:LiNbO$_3$ waveguide \cite{Wang:18}.

In this work, we demonstrate a TFLN quantum pulse gate (QPG) with a normalized conversion efficiency three orders of magnitude higher than that of prior state-of-the-art devices, while preserving all other key performance metrics. We lift the restriction on operation wavelengths and polarizations imposed by bulk material dispersion through engineering the waveguide geometry. The resulting modal dispersion enables group velocity matching of signal and pump in a type-0 process allowing us to confirm high quality QPG operation with an average temporal-mode selectivity of $(96.8 \pm 1.7)\%$ on par with existing demonstrations. 
Crucially, the synergy between the type-0 interaction and strong optical confinement yields an internal normalized conversion efficiency of $(1810 \pm 10)\,\mathrm{W}^{-1}\mathrm{cm}^{-2}$, a conservative lower bound that represents a 1000-fold improvement over previous demonstrations \cite{gil2021improved}.
This exceptional performance facilitates operation in the high-gain regime using accessible pump powers. The dependence of conversion efficiency on pump power exhibits distinct signatures of time ordering \cite{reddy2013temporal}, an effect whose systematic study is often hindered by low conversion efficiencies. Our findings thus pave the way for scalable photonic quantum technologies relying on high-performance QPGs.

\section{Dispersion engineering of a QPG in TFLN}

A QPG is based on dispersion engineered SFG, in which an input signal pulse with central frequency $\nu_\mathrm{in}$ interacts with a strong pump pulse with central frequency $\nu_\mathrm{p}$ and is up-converted to a higher-frequency output pulse with central frequency $\nu_\mathrm{out} = \nu_\mathrm{in} + \nu_\mathrm{p}$ \cite{brecht2014demonstration, eckstein2011quantum}. 
In this process, the group velocities of input signal and pump are matched, while the output pulse is typically trailing behind the pump and input. 
The QPG selectively addresses a single TM from the input signal, whose complex-valued spectrum is set by that of the pump pulse, frequency-shifted to the signal's central frequency. 
Classical pulse shaping of the spectral amplitude and phase of the pump pulse then allows to flexibly choose, which signal TM is selected.
To describe the operation principle of a QPG, we use the transfer function in frequency space which determines how input frequencies $\nu_\mathrm{in}$ are mapped to output frequencies $\nu_\mathrm{out}$. The transfer function $G(\nu_\mathrm{in}, \nu_\mathrm{out})$ is a product of the pump function $\alpha(\nu_\mathrm{in},\nu_\mathrm{out})$ which reflects energy conservation and the phase matching (PM) function $\Phi(\nu_\mathrm{in},\nu_\mathrm{out})$ given by the dispersion of the waveguide:
\begin{align}
    G(\nu_\mathrm{in},\nu_\mathrm{out}) = \alpha(\nu_\mathrm{in},\nu_\mathrm{out}) \cdot \Phi(\nu_\mathrm{in},\nu_\mathrm{out}).
\end{align}
\noindent The phase matching function is described by 
\begin{align}
    \Phi(\nu_\mathrm{in},\nu_\mathrm{out}) \propto \operatorname{sinc}\left( \frac{\Delta \beta \cdot L}{2} \right), 
    \label{eq:PMfunction}
\end{align}
where $L$ is the length in which the nonlinear process takes place and $\Delta \beta $ is the phase mismatch between the input, pump and output fields. The waveguide dispersion is chosen to achieve group velocity matching between input signal and pump pulses. Consequently, the phase matching function exhibits an angle of $\phi_{\mathrm{PM}}$=0$\,^\circ$ given by \cite{u2005generation}
\begin{align}
    \phi_\mathrm{\mathrm{PM}} = \mathrm{arctan}\left( \frac{v_{g,\mathrm{p}}^{-1} - v_{g,\mathrm{in}}^{-1}}{v_{g,\mathrm{p}}^{-1} - v_{g,\mathrm{out}}^{-1}} \right),
\end{align}
where $v_{g,i}$ is the group velocity of the field $i$.  Under group-velocity matching, the phase matching function is independent of the input frequencies ($\Phi(\nu_\mathrm{in},\nu_\mathrm{out}) \approx \Phi(\nu_\mathrm{out})$). If the phase matching bandwidth is narrower than the bandwidth of the input and pump pulses, the transfer function becomes separable into the outer product of an input and  an output mode, and the QPG performs a TM selective operation \cite{donohue2018quantum}. In other words, the QPG converts a single TM that is set by the complex-valued spectrum of the pump pulse, while all other TMs are transmitted. We define the temporal-mode selectivity as figure of merit for the QPG operation:
\begin{align}
    \label{eq:mode_selectivity}
    S_i = \frac{|\rho_i|^2}{\sum_{j=1}^{\infty} |\rho_j|^2}.
\end{align}
Here, $|\rho_i|^2$ is the conversion efficiency of the $i^\mathrm{th}$ TM. This quantifies how well the device extracts a target TM from an orthogonal set. A value of $S=1$ indicates perfect selectivity with only the target TM converted, while $S=0$ means that no selection takes place and every input TM is converted with the same efficiency.

\begin{figure}[htpb]
    \centering
    \includegraphics[width=0.95\linewidth]{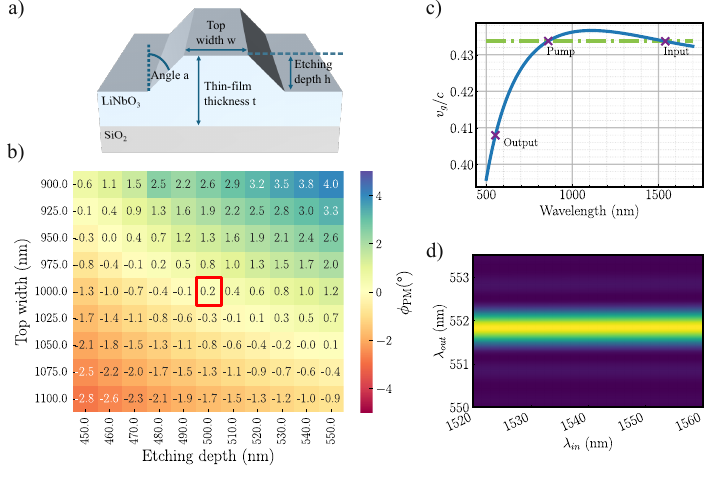}
    \caption{Simulation for finding a suitable waveguide geometry for a TFLN QPG. a) Waveguide geometry in TFLN with the important parameters. b) The heat map shows the phase matching angle $\phi_\mathrm{PM}$ for different top widths $w$ and etching depths $h$. From these simulations we choose the geometry marked with the red rectangle. c) Group velocity as function of the wavelength for the chosen geometry. We mark the output, pump and input signal. The horizontal line shows that this geometry yields group velocity matching between input signal and pump. d) Simulated phase matching intensity $|\Phi(\lambda_\mathrm{in},\lambda_\mathrm{out})|^2$. Due to group velocity matching the phase matching is oriented horizontally.}
    \label{fig:simulation}
\end{figure}

TFLN enables us to harness the strongest nonlinear tensor element $d_{33}=27\,\mathrm{pm/V}$ while remaining compatible with the legacy QPG configuration otherwise. To this end, we identify a waveguide geometry that provides group-velocity matching between an input signal at $\lambda_\mathrm{in}=1540\,\mathrm{nm}$ and and a pump at $\lambda_\mathrm{p}=860\,\mathrm{nm}$ in a type-0 process. In weakly guiding structures such as Ti:LiNbO$_3$ waveguides, the dispersion is governed by the material dispersion, and the only accessible degree of freedom for tuning the group velocities is the birefringence of the crystal. Group-velocity matching between 1540$\,$nm and 860$\,$nm therefore requires the two fields to be orthogonally polarized, which forces the legacy QPG into a type-II process and locks its operating wavelengths to the narrow set of combinations for which the birefringence happens to compensate the material dispersion. In TFLN, by contrast, the high index contrast makes the waveguide dispersion a design parameter in its own right: the geometric contribution can be tailored to compensate the material dispersion for co-polarized fields, so that group-velocity matching becomes achievable in a type-0 process at the same wavelengths. This decouples the choice of process type and operating wavelengths from the material dispersion and is the key enabler for the device presented here. The relevant geometric degrees of freedom are the film thickness $t$, etching depth $h$, top width $w$, and sidewall angle $\alpha$ (see \autoref{fig:simulation}(a)).

To this end, we performed numerical simulations of the effective refractive index for a x-cut, y-propagation 5 mol\% MgO-doped TFLN sample.  The simulations, conducted in Lumerical \cite{Lumerical} using the Sellmeier equations reported in \cite{gayer2008temperature}, consider a fixed geometry defined by a thin-film thickness of $t=600\,\mathrm{nm}$, a sidewall angle of $\alpha=60^\circ$ (constrained by the etching process), and a 2$\,$µm silicon dioxide insulator layer on a silicon handle. We varied the top width $w$ and etching depth $h$ and evaluated the corresponding phase matching angle $\phi_\mathrm{PM}$ (see \autoref{fig:simulation} b). Based on this analysis, we selected a top width of $w=1\,\mu\mathrm{m}$ and an etching depth of $h=500\,\mathrm{nm}$ which yield near-perfect group velocity matching ($\phi_\mathrm{PM}\approx0.15^\circ$). We plot the corresponding group velocities in dependence of the wavelength for the fundamental TE mode in \autoref{fig:simulation} c), where we marked the group velocities of the input, pump and output fields. \autoref{fig:simulation} d) is a plot of the phase-matching intensity $|\Phi(\lambda_\mathrm{in},\lambda_\mathrm{out})|^2$ for a waveguide with a poled length of $L=3\,\mathrm{mm}$ (c.f. \autoref{eq:PMfunction}), which shows the horizontal orientation characteristic for group-velocity matching of input signal and pump. For the optimized waveguide geometry, we estimated an effective mode area of  $A_{\mathrm{eff}} = 0.6\,\mu\mathrm{m}^2$ calculated through  
\begin{align}
A_{\mathrm{eff}} = \prod_{j}
\frac{
\iint |E_j(x,y)|^2 \, dA
}{
\left| \iint 
E_{\mathrm{in}}(x,y)\,
E_{\mathrm{p}}(x,y)\,
E_{\mathrm{out}}^*(x,y)\, dA
\right|^2
},
\end{align}
where $j \in \{\mathrm{in},\mathrm{p},\mathrm{out}\}$ and the spatial distributions $E_j(x, y)$ for the three fields were obtained by using EME  simulations in Lumerical.

Based on the simulations, we fabricated a 1x1$\,$cm TFLN sample (NANOLN) with a measured thin-film thickness of 604$\,$nm, determined via ellipsometry at five different points on the sample. This thickness corresponds to a nominal poling period of 3.178$\,$µm. To mitigate potential fabrication variations, as described in \cite{babel2025ultrabright},  we produce a sample with three waveguide widths (900$\,$nm, 1000$\,$nm, and 1100$\,$nm) and five poling periods (3.008$\,$µm, 3.093$\,$µm, 3.178$\,$µm, 3.274$\,$µm, 3.369$\,$µm). The periodically poled waveguides were fabricated by first applying periodic poling with finger electrodes, followed by etching using inductively coupled plasma reactive ion etching (ICP-RIE) with pure argon ions. Further details on the fabrication process can be found in \cite{babel2025ultrabright}. 

\section{Setup}
\label{sec:setup}
\begin{figure}[t]
    \centering
    \includegraphics[width=0.85\linewidth]{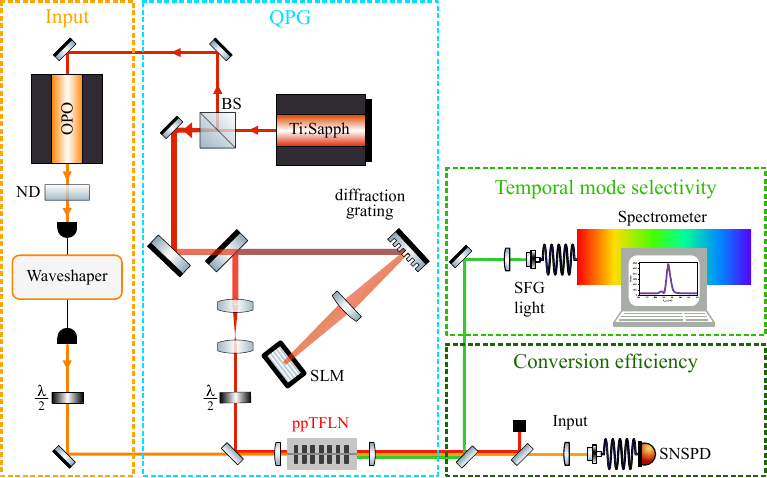}
    \caption{ Sketch of the experimental setup for benchmarking the TFLN QPG. The input signal and pump TMs are shaped using a commercial and a home-built waveshaper, respectively. Both fields are coupled to the sample and the generated green output light is measured with a single-photon sensitive spectrometer while the transmitted input signal photons are counted with superconducting nanowire single photon detectors. BS: beam splitter, $\frac{\lambda}{2}$ : half wave plate, ND: neutral density filter, SLM: spatial light modulator, SNSPD: superconducting nanowire single photon detector. }
    \label{fig:Setup}
\end{figure}

The QPG converts a single TM from an input signal pulse, which is chosen by accordingly setting the TM of the pump pulse.
To verify and benchmark the TFLN QPG performance, we utilize the experimental setup sketched in \autoref{fig:Setup}. 
We generate the pump and input signal pulses with an ultrafast titanium-sapphire oscillator (Ti:Sapph) and an optical parametric oscillator, respectively. The system features an $80\,\mathrm{MHz}$ repetition frequency and is set to central wavelengths of $\lambda_\mathrm{p}=872.6\,\mathrm{nm}$ and $\lambda_\mathrm{in}=1553.4\,\mathrm{nm}$. The input signal, generated by the OPO, is attenuated and sent to a commercial, fiber-coupled waveshaper with a resolution of $5\,\mathrm{GHz}$, that is capable of setting amplitude and phase of each frequency component. The shaped input signal TM is then sent toward the TFLN QPG. The Ti:Sapph pulses are sent through a home-built waveshaper based on a spatial light modulator in a 4f-configuration (see, e.g., \cite{monmayrant2010newcomer}) with a resolution of $10\,\mathrm{GHz}$. After shaping the pump TM, it is sent toward the TFLN QPG, combined with the input signal on a dichroic beam splitter and coupled to the TFLN waveguide using an aspheric lens with a focal length of $f=3.1\,\mathrm{mm}$.

We collimate the output of the TFLN QPG with another lens and separate the remaining pump, transmitted input signal, and converted output with a succession of dichroic beam splitters. We then couple the converted output signal to a single-mode fiber and detect it with a single-photon sensitive spectrometer with a resolution of around $30\,\mathrm{GHz}$. We further couple the transmitted input signal to a single-mode fiber and send it to a superconducting nanowire single photon detector for photon counting.

\section{Measurements and Results}

To benchmark the performance of our TFLN QPG, we measure its phase matching function $\Phi(\nu_\mathrm{in}, \nu_\mathrm{out})$, then assess its temporal-mode selectivity $S$, and finally demonstrate high normalized conversion efficiencies. 

\begin{figure}[t]
    \centering
    \includegraphics[width=0.7\linewidth]{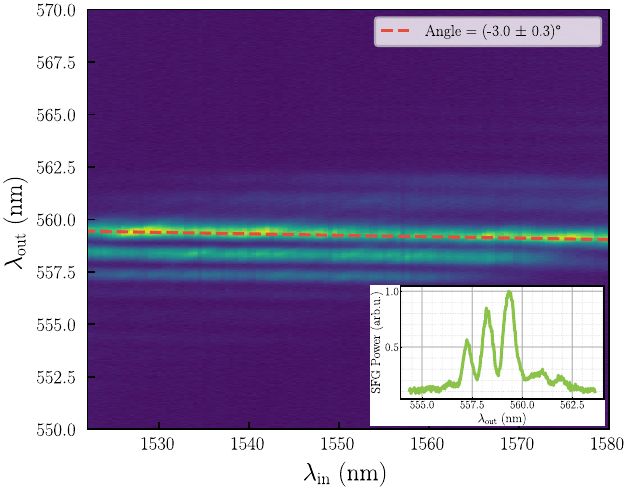}
     \caption{Map of the phase matching intensity $|\Phi(\lambda_\mathrm{in}, \lambda_\mathrm{out})|^2$. The angle of the phase matching function evaluates to $\phi_\mathrm{PM}=(-3.0\pm0.3)^\circ$, thus verifying operation close to perfect group velocity matching. The inset shows a cut through the phase matching intensity at $\lambda_\mathrm{in}=1550\,\mathrm{nm}$. The two prominent side lobes to the left of the main peak are likely caused by fabrication imperfections and will be spectrally filtered during further measurements.}
    \label{fig:measured_phasematching}
\end{figure}

We characterize the phase matching intensity $|\Phi(\nu_\mathrm{in}, \nu_\mathrm{out})|^2$ of the TFLN QPG using a simplified version of the setup in \autoref{fig:Setup}. We do not shape the spectrum of the pump pulses and use a tuneable continuous wave laser as input signal instead of the OPO. We use pump and input powers of $3.5\,$mW and $1.0\,$mW, respectively. First, we fix the central pump wavelengths and scan the tuneable laser on a fine grid. We measure the spectrum of the converted light for each combination of pump and input wavelength and, stitching them together for a set of pump wavelengths, obtain the map of the phase matching intensity in \autoref{fig:measured_phasematching}. We find a phase matching angle $\phi_\mathrm{PM} = (-3.0\pm0.3)^\circ$ which slightly deviates from the expected $0.15^\circ$. The FWHM of the central phase matching peaks is $0.61\,$nm $(583\,\mathrm{GHz})$ in good agreement with the expected 0.627$\,$nm (605$\,$GHz). Finally, the inset in \autoref{fig:measured_phasematching} shows a cut through the phase matching function along the $\lambda_\mathrm{out}$-axis for an input wavelength of $\lambda_\mathrm{in}=1550\,$nm. We see two pronounced side peaks that are blue-shifted from the central peak with the general shape deviating from an ideal sinc-shaped function (see \autoref{eq:PMfunction}). The mismatch in angle and shape indicates a slight imprecision in the simulated Sellmeier equations of our waveguide as well as inhomogeneities in the waveguide geometry that may be caused by fabrication tolerances \cite{santandrea2019fabrication}. We note that, while undesirable, the side peaks in the phase matching do not prevent QPG operation since we can apply spectral filtering to the output to detect only the central peak. In an application, this operation corresponds to added loss on the output light of the QPG which can be alleviated through future optimization of fabrication processes.

Next, we assess the temporal-mode selectivity $S$ of the TFLN QPG using the setup in \autoref{fig:Setup}. For this, we need to measure the conversion efficiency of different TMs from an orthogonal set for any given target TM. We choose Hermite-Gauss (HG) TMs of orders 0 to 4 for benchmarking the TFLN QPG operation to guarantee comparability with previous implementations based on Ti:LiNbO$_3$ waveguide \cite{ansari2017temporal,serino2023realization}. We set a spectral-amplitude FWHM of $1.0\,$THz which is larger than the phase matching bandwidth as per the requirement for high-quality operation. We set the target TM by shaping the spectral amplitude and phase of the pump to the $j$-th HG mode. We then cycle the input signal TM through all five HG modes of the set and measure the respective output counts. To this end, we record the output spectrum with a single-photon sensitive spectrometer, subtract the background (measured without any light in the TFLN QPG), and apply a spectral filter with a width of $0.1\,$THz around the center of the main peak of the phase matching to reduce the impact of the side peaks (c.f. \cite{ansari2017temporal,ansari2018tomography,serino2023realization}). We then integrate over the remaining area to obtain the number of converted counts. 

\begin{figure}[t]
    \centering
    \begin{subfigure}{0.495\textwidth}
    \raggedright a)\\
        \centering
        \includegraphics[width=\textwidth]{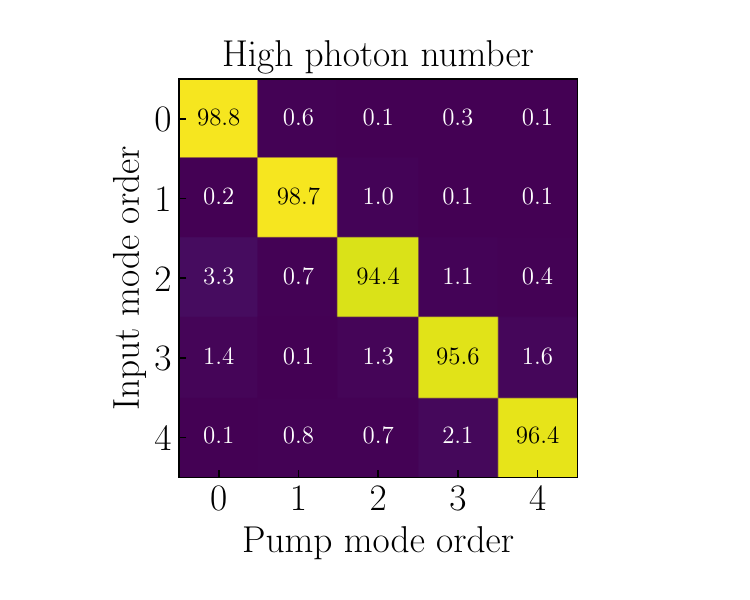}
        \label{fig:bild1}
    \end{subfigure}
    \begin{subfigure}{0.495\textwidth}
    \raggedright b)\\
        \centering
        \includegraphics[width=\textwidth]{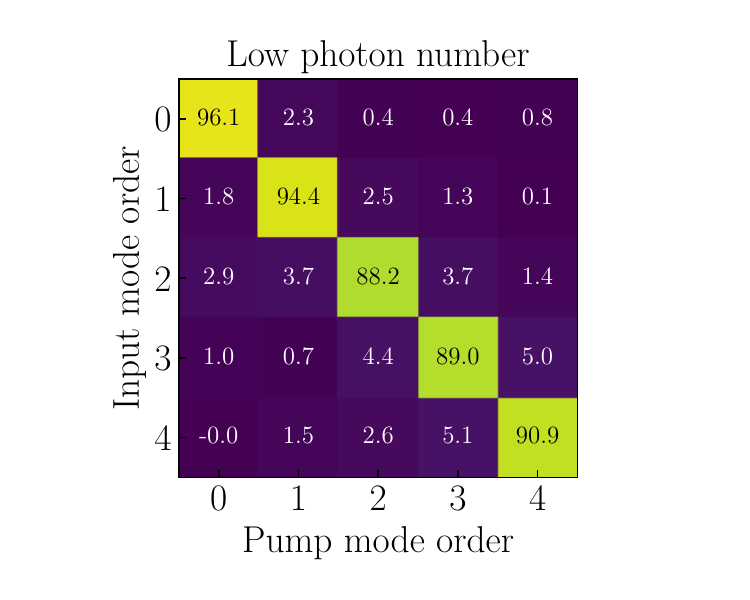}
        \label{fig:bild2}
    \end{subfigure}
     \caption{Temporal-mode selectivity plots for a) high and b) low input signal photon numbers. The numbers are the percentage of converted counts of a specific input signal TM for different pump TMs. The average temporal-mode selectivity for high photon numbers of around $3\cdot10^4$ per pulse is $(96.8\pm1.7)\%$, while it slightly reduces to $(91.7\pm3.1)\%$ for an average photon number of one per pulse, likely due to measurement noise.}
    \label{fig:selectivity}
\end{figure}

\autoref{fig:selectivity} a) shows the results for pump and input powers of $2.0\,$mW and $0.3\,$µW (corresponding to around $3\cdot10^4$ photons per input pulse) in front of the waveguide, respectively. The dominant diagonal means that conversion primarily takes place when the input TM matches the target TM of the TFLN QPG and we obtain a high average temporal-mode selectivity of $\bar{S}=(96.8\pm1.7)\%$ over all target TMs. Next, we reduce the input photon number to, on average, one photon per pulse to demonstrate single photon operation. This step is crucial in QPG benchmarking because pump-induced noise can inhibit single-photon operation \cite{reddy2018high} and would show up as a steep drop in $\bar{S}$. The results of this measurement, shown in \autoref{fig:selectivity} b), yield $\bar{S}=(91.7\pm3.1)\%$, which is limited by a reduced signal to noise ratio due to detection noise. 
Our measurements confirm that the TFLN QPG exhibits an excellent temporal-mode selectivity on par with previous QPGs in Ti:LiNbO$_3$ waveguides \cite{brecht2014demonstration,ansari2017temporal,serino2023realization}. Note that all measurements of temporal-mode selectivity were measured in the low-gain regime.

Finally, we measure the internal conversion efficiency of our TFLN QPG as function of the pump power for an input average photon number of 100 photons per pulse. For this measurement, both pump and input are set to the $0$-th order HG TM, which showed the best temporal-mode selectivity, see \autoref{fig:selectivity}. We now measure the transmitted input signal using superconducting nanowire single-photon detectors (SNSPD) instead of the converted output. This measurement mode allows for evaluating the conversion efficiency inside the TFLN QPG by measuring the depletion of the input signal while being immune to additional external losses on the converted light as caused by spectral filtering of the phase matching or limited fiber coupling efficiencies.

We formally define the internal conversion efficiency as (c.f. \cite{allgaier2017highly})
\begin{align}
    \eta = 1 - \frac{C_{\text{pump on}}}{C_{\text{pump off}}}, \label{eq:efficiency}
\end{align}
where $C_{\text{pump off}}$ are the measured counts when the input signal is coupled and transmitted through the waveguide and the pump is blocked, and $C_{\text{pump on}}$ are the measured counts when the pump is open and the conversion process takes place. We record photon counts over a $2.6\,$ns temporal window using a time tagger with a resolution of $1.0\,$ps. From this time trace, we obtain the total counts through integration over a $120\,$ps window centered on the arrival time peak of the input signal. The window was chosen to be larger than the combined input signal pulse duration and detection timing jitter, while still removing as many background noise counts as possible. For each run of the experiment, we measured background noise counts by blocking the input signal while leaving the pump on. These were subtracted from the measured counts. The reported counts correspond to the mean of ten acquisitions per pump power while the uncertainties represent one standard deviation. 

\begin{figure}[t]
    \centering
    \includegraphics[width=0.7\linewidth]{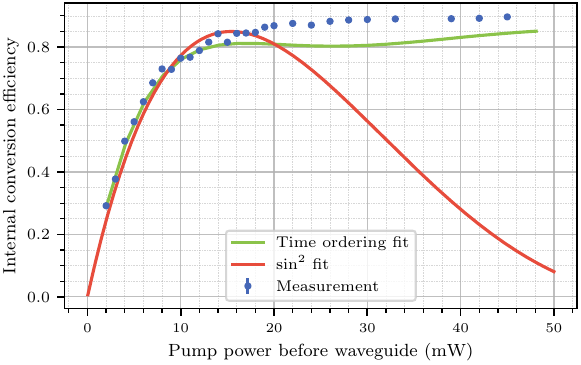}
    \caption{Internal conversion efficiency as a function of the pump power in front of the waveguide. Blue dots are the measurement, error bars are smaller than the symbol size. The red line is the sine-squared function. The green line is the prediction of a numerical model that captures time-ordering effects in the process. The model reproduces the measurement well for lower pump powers. Deviations at higher pump powers may originate from parasitic nonlinear effects and spectral pulse phases that are not captured in the model.}
    \label{fig:efficiency}
\end{figure}  

\autoref{fig:efficiency} shows the results of this measurement. We achieve a conversion efficiency of around $89\%$ for a pump power of around $20\,$mW. 
Following \cite{Fejer2004SFG}, we fit a sine-squared function (red line) to the low-conversion efficiency ($<40\%$) region of the measurement. From this, we obtain a normalized conversion efficiency of $\eta_\mathrm{norm}=(1810\pm10)\,$W$^{-1}$cm$^{-2}$.
This value is a strict lower bound, since all pump powers are measured in front of the waveguide. Simulations predict a maximum coupling efficiency of 30\% for the pump, so the power actually available inside the waveguide is at least a factor of three lower than plotted, and $\eta_\mathrm{norm}$ correspondingly at least a factor of three higher than quoted. We nevertheless report the uncorrected value in order to keep the comparison with previous work conservative, and note that the systematic uncertainty of $\eta_\mathrm{norm}$ is dominated by this coupling estimate rather than by the fit.
We emphasize that even this conservative lower-bound estimate exceeds the normalized conversion efficiency reported for existing QPGs in Ti:LiNbO$_3$ waveguides by three orders of magnitude \cite{gil2021improved}.

Theoretical scaling arguments predict that the normalized conversion efficiency should increase by approximately four orders of magnitude compared to prior state-of-the-art devices. This enhancement stems from two primary factors:
(i) the transition from a type-II to a type-0 process ($d_{33}=27\,\mathrm{pm/V}$ vs $d_{31}=4.5\,\mathrm{pm/V}$, and $\eta\propto d_{xy}^2$), and
(ii) the significantly stronger optical confinement in TFLN waveguides ($A_\mathrm{eff, TFLN}\approx0.6\,\mu\mathrm{m}^2$ vs $A_{\mathrm{eff, Ti:LiNbO}_3}\approx 64\,\mu\mathrm{m}^2$).
Our measured efficiency is roughly one order of magnitude below this ideal prediction, a discrepancy we attribute to conservative estimates of the effective pump power, consistent with expected pump coupling losses of $\sim 10\%$, and deviations from ideal phase matching. Accounting for these factors brings our experimental results into agreement with theoretical expectations.

Finally, we analyze the conversion efficiency curve in \autoref{fig:efficiency}. We observe a significant deviation from the standard sinusoidal behavior typically assumed for extracting the normalized conversion efficiency. This behavior serves as a strong indicator for the impact of time-ordering effects in the TFLN QPG \cite{reddy2013temporal,christ2013theory}. A numerical simulation incorporating these full temporal dynamics reproduces the experimental trend (green line) in good agreement. Minor discrepancies at high pump powers may arise from pulse chirp or nonlinear dispersion not included in the model. These results confirm that QPG operation in the high-gain regime is strongly influenced by time-ordering, leading to substantial modifications of the temporal eigenmodes during propagation. While theory predicts a concomitant decrease in temporal-mode selectivity \cite{reddy2013temporal}, an experimental verification of this prediction remains a target for future work.

\section{Conclusion}

In this work, we have demonstrated the first TFLN QPG, leveraging the platform’s high normalized conversion efficiency and broad flexibility in operating wavelengths and polarization configurations to enable efficient and versatile QPG operation.
We designed a waveguide geometry with tailored dispersion that supports group-velocity matching of the input signal and pump fields in an efficient type-0 SFG process. 
After fabrication, we experimentally verified a phase matching angle of $\phi_\mathrm{PM} = (-3.0 \pm 0.3)^\circ$, which is close to perfect group velocity matching. 
This demonstrates that the TFLN QPG overcomes one of the two limitations of traditional QPGs: 
restricted operating wavelengths and polarizations. 
To verify QPG operation, we measured an average TM-selectivity of $\bar{S}=(96.8 \pm 1.7)\%$ for bright input signals and $\bar{S}=(91.7 \pm 3.1)\%$ for single-photon level input signals. 
These numbers verify the high quality operation of our TFLN QPG and put it on par with earlier QPGs in Ti:LiNbO$_3$ waveguide. 
Finally, we demonstrate an exceptional internal conversion efficiency of $\eta=(89.6 \pm 0.1)\%$ for a pump power of only $20\,\mathrm{mW}$ in front of the waveguide. 
From this we extract a conservative lower-bound estimate of the normalized conversione efficiency of $\eta_\mathrm{norm}=(1810 \pm 10)\,\mathrm{W}^{-1}\mathrm{cm}^{-2}$, with true values up to an order of magnitude higher due to an estimated incoupling efficiency on the order of 10\% for the pump field. 
This outperforms existing QPGs by at least three orders of magnitude. 
Thus, the TFLN QPG also provides a solution to the second limitation of traditional QPGs: 
the limited normalized conversion efficiency. 
On a different note, the conversion efficiency curves also indicate that, as a result of the high efficiency of the TFLN QPG, the high-gain regime in which time-ordering effects play a significant role can easily be reached. 

Our results exemplify that TFLN is an ideal platform for realizing high-performance QPGs, offering simultaneously enhanced efficiency, flexible wavelength combinations through advanced dispersion engineering, and increased integration density through miniaturization. 
While further improvements in coupling efficiencies are necessary for practical applications that go beyond laboratory demonstrations, this work lays the foundation for complex quantum experiments based on QPGs previously limited by experimentally achievable conversion efficiencies. 
Examples include high-dimensional quantum key distribution with improved secret key rates that can be adapted to different network topologies in realtime \cite{serino2025programmable}, or large frequency encoded quantum networks \cite{folge2024framework} with high-efficiency operations that provide advanced state engineering and overall success rates. 
Additionally, these results bring into reach multi-stage QPG architectures that leverage the high integration density provided by TFLN to mitigate the adverse effects of time-ordering on the QPG operation \cite{reddy2018high}. 
This then paves a realistic way towards applications that rely on high conversion efficiencies for optimal performance such as coherent noise filtering for communication in the photon starved regime \cite{raymer2020time} or efficient frequency interfaces for hybrid quantum networks that heterogeneously combine different kinds of sources, emitters, and quantum memories. 
Ultimately, these results also underscore the significant impact that TFLN may have on quantum photonic technologies.

\begin{backmatter}
\bmsection{Funding}
Deutsche Forschungsgemeinschaft (DFG, German Research Foundation) – SFB-Geschäftszeichen TRR142/3-2022 – Projektnummer 231447078; Max Planck School of Photonics. This work was supported by the European Union's Horizon Europe research and innovation programme through the Marie Skłodowska-Curie Doctoral Network MICROCOMBSYS under grant agreement No. 101119968 and by the NSF Engineering Research Center for Quantum Networks No. EEC-1941583 and NSF QuIC-TAQS under Award No. 2138068.

\bmsection{Acknowledgments}
Silia Babel is part of the Max Planck School of Photonics supported by the Dieter Schwarz Foundation, the German Federal Ministry of Research, Technology and Space (BMFTR), and the Max Planck Society.  Alejandra Alarcón is supported by the H2020 Marie Skodowska Curie Innovative Training Network Microcombsys.
We thank René Pollmann, Franz Roeder, Fabian Schlue and Jonas Babai-Hemati for helpful discussions.
Portions of this manuscript were drafted with the assistance of AI. The final text was reviewed and edited by the authors.

\bmsection{Disclosures}
Marko Lončar is involved in developing lithium niobate technologies at HyperLight Corporation. The remaining authors declare no competing interests.

\bmsection{Data Availability Statement}
Data underlying the results presented in this paper are not publicly available at this time but may be obtained from the authors upon reasonable request.

\end{backmatter}

%%%%%%%%%% If using BibTeX:
\bibliography{sample}

\end{document}